\documentclass[11pt,a4paper]{article}
\usepackage{amsmath,amsfonts,amssymb,amsbsy,bbold,mathtools}
\usepackage{mathrsfs,latexsym,tikz}
\usepackage{slashed}
\usepackage{graphicx}
\usepackage{subcaption}
\usepackage{color}
\usepackage{booktabs}
\usepackage{multirow}
\usepackage{multicol}
\usepackage{alpha}
\usepackage{mathrsfs}

\usepackage{textcomp}
\usepackage{pdflscape}

\usepackage{macros_alpha}

\usepackage{listings}

\lstdefinelanguage{modelFile}{
   morekeywords={Flavours:, Transf:, Discrete:, Spurion:, Block:, Op:},
   comment=[l]{\#},
   alsodigit={:}
}
\AtBeginDocument{\DeclareCaptionSubType{lstlisting}}

\usepackage[normalem]{ulem}

\def\ord{\mathop{\mathrm{O}}}

\input{defs.sty}

\begin{document}

\preprintno{%
CERN-TH-2025-223\\
IFT-UAM/CSIC-25-138
}

\title{opbasis -- a Python package to derive minimal operator bases}

\author[cern,ift,uam]{Nikolai~Husung}
\address[cern]{Theoretical Physics Department, CERN, 1211 Geneva 23, Switzerland}
\address[ift]{Instituto de Física Teórica UAM-CSIC, C/ Nicolás Cabrera 13-15, Universidad Autónoma de Madrid, Cantoblanco 28049 Madrid, Spain}
\address[uam]{Departamento de Física Teórica, Universidad Autónoma de Madrid, Cantoblanco 28049 Madrid, Spain}

\vspace*{-1cm}

\begin{abstract}
Finding a complete and yet minimal on-shell basis of operators of a given mass-dimension that are compatible with a specific set of transformation properties is the first step in any Effective Field Theory description.
This step is the main bottleneck for systematic studies of leading logarithmic corrections to integer-power lattice artifacts in Symanzik Effective Field Theory targeting various local fields and lattice actions.
The focus on discrete symmetry transformations in lattice field theory, especially reduced hypercubic spacetime symmetry with Euclidean signature, complicates the use of standard continuum field theory tools.

Here, a new Python package is being presented that targets the typical lattice field-theorist's use cases.
While the main target lies on continuum EFTs describing 4D non-Abelian lattice gauge theories, the applicability can be extended beyond Effective Field Theories.
New discrete symmetries, twisted masses, or the introduction of boosts are just a few examples of possible extensions that can be easily implemented by the user.
This should allow for a wider range of theories and applications beyond the initial focus of this package.

The general functionality of the package is explained along the lines of three examples: The $\ord(a)$ operator basis of the axial-vector in Wilson QCD, operator bases compatible with the symmetries of unrooted Staggered quarks as well as a pedestrian derivation of a $B^*(\mathbf{p})\pi(-\mathbf{p})$ operator with pseudo-scalar quantum numbers.
Each example makes use of an increasing range of features and requires user-defined extensions show-casing the versatility of the package.
\end{abstract}


\maketitle

\section{Introduction}
The focus of the Python\footnote{\url{https://www.python.org/}} package presented here lies on non-Abelian Euclidean Lattice Gauge Theories with a single coupling constant and traceless generators. 
By introducing a lattice spacing as the UV regulator one trades continuous rotations for discrete ones -- commonly on a hypercubic lattice but other geometries should be accessible too.
Tools commonly used for continuum Effective Field Theories~(EFT) typically have Minkowskian signature with Lorentz symmetry in four space-time dimensions build in or even target directly dimensional regularisation with non-integer dimensions.
An adaptation of such tools to the cases we are interested in would then amount to a major overhaul.
Thus motivating the development of this package.
The package is made available via the repository \url{https://github.com/nikolai-husung/opbasis} under the MIT license.
The paper discusses the current version~0.9.7~\cite{Husung:opbasis}.
Any future changes that affect the workflow of the package will be highlighted in the change notes.

Both EFTs and Renormalisation rely heavily on symmetry constraints and transformation properties limiting the set of operators allowed to contribute.
In the language of EFTs those (higher-dimensional) operators amount to corrections in the local effective Lagrangian or to local composite fields from physics at and beyond some energy cut-off.
An example frequently encountered in the lattice community is Symanzik EFT~(SymEFT) \cite{Symanzik:1979ph,Symanzik:1981hc,Symanzik:1983dc,Symanzik:1983gh} describing lattice artifacts, which we will focus here on.
For Renormalisation on the lattice, symmetry constraints play an even bigger role.
Due to having a dimensionful cut-off, power-divergences are a more severe problem and must be removed by additive renormalisation.
Moreover, lattice regularisations frequently sacrifice some symmetries such as flavour~\cite{Wilson:1974,Wilson:1975id,Kogut:1974ag}, Euclidean reflections~\cite{Kogut:1974ag} etc.~to deal with the Nielsen-Ninomiya no-go theorem~\cite{Nielsen:1980rz,Nielsen:1981xu} as well as to realise cheaper~\cite{Kogut:1974ag} or more stable numerical setups~\cite{Frezzotti:1999vv,Frezzotti:2000nk}.
In turn this leads to more complicated patterns in operator mixing under renormalisation, see e.g.~\cite{Papinutto:2016xpq}, and frequently requires the restoration of Ward identities broken by the chosen regularisation, such as the PCAC Ward identity for Wilson quarks by additive renormalisation of the quark masses as well as finite renormalisation of the axial vector see e.g.~\cite{Luscher:1996ug}.

While having an understanding of the realised symmetries and transformation properties is an important ingredient, working out all operators allowed by those constraints is still a daunting task, especially once one aims at operators with increasing mass-dimensions.
If one further aims at a minimal operator basis after applying equations of motions~(EOM) and possibly identifying total derivatives this quickly becomes a very tedious task and requires strict book-keeping.
In particular for a systematic SymEFT analysis of the asymptotically leading lattice artifacts, see e.g.~\cite{Sheikholeslami:1985ij,Balog:2009np,Husung:2022kvi,Husung:2024cgc}, this is the major bottleneck if one is interested in different lattice discretisations and various local composite fields.
Each of them adding their own distinct symmetries or transformation properties respectively.

This Python package allows to first obtain an overcomplete basis compatible with the desired transformation properties and, subsequently, to reduce it to a minimal basis by working out linear dependencies among the various operators.
The basic concepts of the package are discussed in \sect{sec:concepts}.
To allow for a less abstract discussion of the package, we will rederive along the lines the minimal operator basis for the axial-vector of a isospin doublet to $\ord(a)$ in Wilson QCD, cf.~\cite{Luscher:1996sc}, from the definition of a model in \sect{sec:modelDef} over finding an overcomplete basis in \sect{sec:overcompleteBasis} to the reduction to a minimal basis in \sect{sec:basisReduction}.
Gradually more involved examples will then be used to outline more advanced features in \sect{sec:examples}.
Eventually, we work out the basis for the $\ord(a^2)$ lattice artifacts of unrooted Staggered quarks~\cite{Kogut:1974ag} and a suitable $B^*\pi$-interpolator with pseudo-scalar quantum numbers in \sects{sec:exampleStaggered} and \ref{sec:exampleBstarPi} respectively.

\section{Basic concepts}\label{sec:concepts}
Each operator of interest here is assumed to be scalar but may otherwise transform with a phase of modulus 1.
Such an operator can therefore be written as a product of traces in colour space and bilinears in spinor space.
We only allow for four-fermion operators of the two forms
\begin{equation*}
(\bar{p}\ldots q)(\bar{r}\ldots s),\quad(\bar{p}\ldots T^a\ldots q)(\bar{r}\ldots T^a\ldots s)
\end{equation*}
with fermion flavours $p$, $q$, $r$, $s$ and all indices contracted within each set of brackets except for the colour-algebra index $a$.
This requires the use of generalised Fierz identities by the user to always ensure this structure especially when introducing explicit generators of a discrete flavour symmetry.
For a consistent operator basis all operators considered must be below the mass-dimension of a six-fermion operator (if the theory contains fermions), i.e., below mass-dimension~$9$ in a four-dimensional theory.
This should not be a relevant limitation, for typical scenarios in lattice field theory.\footnote{Relieving this limitation would require the implementation of explicit indices, which would complicate the setup considerably.
Alternatively, writing out explicitly those additional indices would significantly enlarge the expressions representing the operators.}

The overall strategy implemented in the Python package can be summarised as follows:
\begin{enumerate}
\item[1.] Define a model by declaring all discrete transformations realised by the theory, e.g., discrete rotations, reflections etc., any spurionic symmetries, and the flavour content (if any).
This step also requires the implementation of any ``non-standard'' transformations, i.e., transformations beyond the standard hypercubic symmetry transformations or modifications thereof.
\item[2.] Obtain templates classifying all the operator candidates at a given canonical mass-dimension.
A template still contains placeholders for indices, Gamma structures etc.
There are two strategies available to propose these templates, allowing a higher degree of automation versus a more fine-grained control.
\begin{enumerate}
\item[(a)] Get all templates at a specific mass-dimension while tracking the specifics like flavours, number of total derivatives etc.~to allow for manual filtering and / or sorting of the templates.
Still requires explicit choice of flavours to be used in bilinears if any.
\item[(b)] Specify the templates by hand.
This can be useful for more advanced applications of the package or when one solely wants to know all permissable operators of a specific class of operators.
\end{enumerate}
Sorting of the templates will become relevant for the deduction of the minimal basis.
To enforce use of zeroes due to EOMs or integration-by-parts relations, EOM-vanishing operators and total derivatives have to be sorted to the top of the list respectively.
Keep in mind that any manual filtering of the full field content may lead to an incompleteness of the operator basis.
\item[3.] Writing out the templates obtained in step 2 into all possible combinations of indices, Gamma structures and so on yields the building blocks for the future operator basis.
Given the transformation properties specified for the current model, we can then use the standard projection formula \eq{eq:transfProj} for cyclic groups to construct valid operators from such a building block.
Incompatible building blocks automatically project to zero.
For more details see the discussion in \sect{sec:overcompleteBasis}.
\item[4.] Eventually, the reduction to a minimal basis can be achieved iteratively by keeping only those operators that are linearly independent from the ones already added to the minimal basis.
By sorting the templates and thus the operator candidates derived from those templates, we automatically make use of any EOMs or integration-by-parts identities if we keep the ordering intact.
This ordering can be adjusted according to the basis being derived.
\end{enumerate}

\subsection{Python representation of an operator}
Within the Python package an operator \op{} is represented by an instance of \LC{} as visualised in~\fig{fig:LinearComb}.
As the name suggests, \LC{} handles linear combinations of the various terms forming the operator and is able to simplify the linear combination.
This simplification can be enforced by a call to \texttt{op.simplify()} to take care of joining terms differing solely in their prefactors, identifying overall prefactors, imposing a unique ordering among different terms etc.
After simplification the \LC{} has a unique string representation accessible via \texttt{str(op)}.
Prior to calling \texttt{str(op)}, the user has to make sure that \op{} is indeed simplified or otherwise call \texttt{op.simplify()}.
Making this the user's responsibility is a question of efficiency to only perform this step when it is required.
If the simplification step is omitted \texttt{str(op)} may not be used as a unique identifier and some internal \texttt{assert} may even fail entirely.

Each term in \LC{} consists of an overall prefactor and a product of various instances of \Bilin{} and \Tr{}.
The general structure of \Bilin{} and \Tr{} is being visualised in \figs{fig:Bilin}, \ref{fig:colourTrace}, and \ref{fig:otherTrace} with some additional remarks on covariant derivatives, EOMs etc.
Two variants of \Tr{} are implemented.
The more common variant is a \Tr{} in colour-space which may \emph{only} consist of instances of the protected \AlgBlock{} as shown in \fig{fig:colourTrace} and does not allow for any customisation.
A second variant of \Tr{} as visualised in \fig{fig:otherTrace} is intended for traces of quark-mass matrices but may also contain custom implementations of the base-class \Block{} such as, e.g., quark-mass mistunings when working with chirally twisted Wilson quarks.
No simplification is implemented for \Tr{} other than use of cyclicity to ensure unique string representations out of the box.

\subsection{Defining a model}\label{sec:modelDef}
Before we can make use of the package we first need to declare what model we are considering, i.e., transformations, flavour content etc.
For our example of the axial-vector in the isospin basis
\begin{equation}
\AV_\mu^a=\bar{\Psi}\gamma_\mu\gamma_5\tau^a\Psi
\end{equation}
in Wilson QCD, such a model file then looks like listing~\ref{code:axialModelTransf}.
For a brief recap of Wilson QCD see \app{sec:Wilson}.
\begin{figure}
\begin{lstlisting}[language=modelFile,caption={Axial-vector model file declaring transformation properties for the SymEFT basis.},label=code:axialModelTransf]
Transf: axial         # Script holding transformations.
Flavours: up dn = Psi # Flavour content and flavour vector.
Discrete: P +---      # Euclidean reflections.
Discrete: C +         # Charge conjugation.
Discrete: H4 xxx+++   # Hypercubic rotations.
Discrete: tau +--     # (Discrete) SU(2) flavour rotations.
\end{lstlisting}
\end{figure}
The crucial part is the declaration of the various \Discrete{} transformations corresponding to our composite field of interest --- we choose $\mu=0$ and $a=1$ but other choices are easily accessible by adapting the transformation behaviour accordingly.

Each \Discrete{} transformation has to be given a unique name as well as the desired transformation property, where the user can choose from \texttt{+} or \texttt{-} corresponding to $+1$ or $-1$ respectively and \texttt{x} if a transformation is to be omitted.
Supplying multiple characters amounts to defining different directions of the transformation,\footnote{Internal numbering starts at 0.} e.g., for hypercubic rotations one has the six planes $(\mu,\nu)=(0,1),(0,2),(0,3),(1,2),(1,3),(2,3)$.
Eventually, the details of the various transformations have to be specified by the user in \texttt{axial.py}, where each \Discrete{} transformation has a corresponding Python implementation with identical name.

In our first example we only need to implement the transformation behaviour under discrete flavour rotations by $\tau^j$ with $j=1,2,3$, here labelled \texttt{SU2} to pick out the desired element of the SU(2) algebra as highlighted in listing~\ref{code:axialSU2rotation}.
The Pauli matrices themselves are already part of the \texttt{opbasis} package.
All other transformations like reflections or rotations are already built-in and implementing them thus amounts to returning the default behaviour made available in \texttt{default.py}.
\begin{figure}
\begin{lstlisting}[language=Python,caption={Part of \texttt{axial.py} implementing discrete $\tau^j$ flavour rotations.},label=code:axialSU2rotation,firstline=3,firstnumber=3,lastline=7]
def tau(op:opb.LinearComb, j:int):
   tj = opb.SU2(opb.Pauli(j+2))
   for b in op.bilinears:
      b.blocks = [tj] + b.blocks + [tj]
   return op
\end{lstlisting}
\end{figure}

\subsection{Finding an overcomplete basis}\label{sec:overcompleteBasis}
With the definition of the model at hand, we now can turn our attention to which operators we are actually interested in, e.g., bilinears or gluonic traces, total derivatives or EOM-vanishing ones and which mass-dimension these operators should have.
All of this is controlled by so called \emph{templates}.
While those can in principle be provided by the user, it is recommended to rely on a call to \texttt{getTemplates(mdTarget,...)}, which then returns all templates with mass-dimension \texttt{mdTarget}.
Various optional arguments can be used to control which templates are actually generated including a way to filter sub-templates through a user-defined function handle.
Obviously, missing any templates in this step may invalidate the completeness of the final operator basis.
The templates being returned are still carrying some metadata like the number of masses, flavour content, derivatives etc. that can then be used for customised ordering of the templates or even posterior filtering.
Once all preprocessing is finished one can obtain the string-representation of each template via \texttt{str(...)}.

Here we are looking for operators of mass-dimension~4, that are not EOM-vanishing and have the appropriate transformation properties to describe the $\ord(a)$ lattice artifacts of the chosen axial-vector.
To achieve this we need to have all relevant operators \emph{including} EOM-vanishing ones to then infer the minimal on-shell basis linearly independent from the latter variants and among each other.
How linear-independence is worked out in practice will be explained in some detail in \sect{sec:basisReduction}.
The inclusion of total derivatives and EOM-vanishing operators is realised by templates carrying an explicit overall \texttt{d} for a total derivative and \texttt{D0l} or \texttt{D0} for insertions of the fermion EOM acting to the left or right respectively.
Internally the canonical transformations for \texttt{D0} and \texttt{D0l} under Euclidean reflections, rotations etc.~are implemented and the user has to adjust the behaviour if there is a non-trivial modification needed.

From the templates we may now generate a set of ans\"atze that includes all possible combinations of (generalised) indices, such as Lorentz indices, different Dirac-Matrices, and so on.
The derivation of an overcomplete basis relies on the fact that all transformations considered become the identity after $n$ iterations up to an overall sign, i.e.,
\begin{equation}
\opm=\opm^{[0]}\rightarrow \kappa\opm^{[1]}\rightarrow \ldots\rightarrow \kappa^n\opm^{[n]}\equiv\pm\opm\,,
\end{equation}
where $\kappa=\pm 1$ are the user-defined transformation properties.\footnote{Conceptually, the generalisation to $\kappa$ being a complex phase with rational multiple of $\pi$ should be feasible but is beyond the scope of this package.}
In case an ansatz is indeed a building block for an invariant operator, $n>0$ is the minimal number of applications of the transformation to come back to the original ansatz, e.g., for reflections, $n=1$.
If the ansatz is not a valid building block, only $2n$ iterations will suffice and the $n$th iteration is instead identical to $-\opm$.
This fact can then be used by applying each transformation $n'$ times until the ansatz is recovered including the sign.
Then all $n'-1$ steps are summed up, i.e.
\begin{equation}
\opm_\mathrm{new}\propto\sum_{l=0}^{n'-1}\kappa^l\opm^{[l]}\,.\label{eq:transfProj}
\end{equation}
If the ansatz was a compatible building block this sum will now be invariant under each separate iteration of the transformation, while each incompatible ansatz will sum to zero and can be discarded.
Here, this procedure is being generalised further for transformations that have multiple directions such as rotational planes by performing the same iteration for all directions on each newly found building block until no new building blocks arise.
Afterwards all unique building blocks are being summed analogously to \eq{eq:transfProj}.
Performing the same iterative step on the various transformations results in a set of $\opm_\mathrm{new}$ that indeed have the desired transformation properties.
Since the initial ans\"atze are severely overcomplete, we only keep those $\opm_\mathrm{new}$ that differ in their string-representation \texttt{str(opnew)} disregarding overall prefactors.
This finally yields a basis of operators that is complete (if and only if the templates provided are complete) and minimal up to the use of EOMs, integration by parts relations or identities like $F_{\mu\nu}=[D_\mu,D_\nu]$.

In practice all of this is implemented by a call to \texttt{overcompleteBasis(template, model)}, which takes as arguments a String describing the current \texttt{template} as well as the Python representation of the \texttt{model} definition and eventually returns the list of operators remaining after applying the procedure described above.

\def\linesEOMsAxial{51--70}
\subsection{Reduction of the operator basis}\label{sec:basisReduction}
Which basis one is actually looking for depends a lot on the problem at hand.
For example, for local fields one has to keep total derivatives in the minimal basis or, if one is interested in an off-shell renormalisation prescription, one needs to keep EOM-vanishing operators.
In the latter case one may still want to distinguish between the ``on-shell-ness''.\footnote{The package does not target (directly) perturbation theory and thus does not care for gauge-fixing, ghosts etc.}

For identifying linear dependencies among all the operators in our overcomplete basis we first need to establish a common (and truly unique) representation.
This includes writing out the placeholders of any insertions of the EOMs, which have to be specified at this point --- for our example this is done in lines \linesEOMsAxial{} in \texttt{findBasisAxial.py} as part of the supplemental material.
Further we write out all \LC{} as follows:
\begin{enumerate}
\item By introducing total derivatives we can ensure that within each \Bilin{} all covariant derivatives act to the right.
\item Write out all covariant derivatives and field-strength tensors explicitly in terms of derivatives and gauge fields (and derivatives acting on gauge fields).
\item For any instance of a colour-space \Tr{}, all total derivatives are written out explicitly acting on the elements in the trace and then cyclicity is used to arrive at a unique ordering.
\end{enumerate}
Once each \LC{} has been reexpressed internally, the associated String representation of \emph{each term} is truly unique.
This allows us to map each \LC{} into a vector $v_j$ in the vector-space $V$ of all the terms within the overcomplete basis.
On this vector-space we can then enforce linear-independence to find the minimal basis of the set of vectors $\{v_j\}_{j\in\Gamma}$.
Notice that in general $\{v_j\}_{j\in\Gamma}\subsetneq V$.
Here this is implemented as an iterative process as indicated in listing~\ref{code:minimalBasisPseudo}.
\begin{figure}
\begin{lstlisting}[mathescape=true,caption={Pseudo-code highlighting how the minimal operator basis is derived.},label=code:minimalBasisPseudo]
M = $(\mathbb{Q}+i\mathbb{Q})^{\dim(V)\times 0}$
for $j\in\Gamma$
   M' = M
   M'[1:end,end+1] = $v_j$
   if rank(M') > rank(M)
      M = M'
      # add the $j$th operator to the minimal basis
   endif
endfor
\end{lstlisting}
\end{figure}
Somewhat counter-intuitively this approach requires the least relevant operators being added first, namely those operators that establish zeros due to being EOM-vanishing (or total derivatives when being interested in contributions to the action). 
Consequently the ordering in the set of operators $\Gamma$ matters.
This allows the user to choose freely the ordering according to the problem at hand but also according to the user's preferences, e.g., whether to keep total derivatives in favour of operators carrying explicit mass-dependence when aiming at local composite fields.
In our example of the axial-vector we use the following hierarchy
\begin{equation*}
\text{EOM-vanishing} > \text{massive} > \text{total derivatives} > \text{others}\,,
\end{equation*}
which is imposed on the templates by use of listing~\ref{code:axialModelTemp}.
\begin{figure}
\begin{lstlisting}[language=Python,caption={Example how to enforce the desired ordering within the templates for the axial-vector.},firstline=28,firstnumber=28,lastline=36,label=code:axialModelTemp]
templates = [x for x in sorted(templates, reverse=True,
   key=lambda x: (
      + 7**5 * (x["DF"]+x["D0"]+x["D0l"])
      + 7**4 * x["M"]
      + 7**3 * (1 if x.tder else 0)
      + 7**2 * x["Bilinear"]
      + 7    * (1 if "D.F.D.F" in x.rep else 0)
      +        x["F"]*(1 if x["Bilinear"]>0 else -1)
      ))]
\end{lstlisting}
\end{figure}

It should be clear that the user can still reorder within the overcomplete basis.
To avoid reiterating all the preceding steps, it is good practice to store the intermediate basis as is being done in the main script \texttt{findBasisAxial.py}.
The most natural ways of shuffling affect the ordering in which the templates are provided as well as the ordering within a set of operators belonging to a given template.

Eventually we get as output, where the grouping still reflects the ordering among the initial templates,
\begin{lstlisting}[numbers=none,xleftmargin=0em,framexleftmargin=0em,tabsize=3]
+1*{
	+1*<Psi.D0l.Gamma[gw5].SU2[t1].Psi>
	-1*<Psi.Gamma[gw5].SU2[t1].D0.Psi>
}
############################################
+1*{
	+1*<Psi.M.Gamma[gw5].SU2[t1].Psi>
}
############################################
+1*{
	+1*<Psi.Gamma[gw5].SU2[t1].Psi tr(M)>
}
############################################
+1*{
	+1*d[0]<Psi.Gamma[g5].SU2[t1].Psi>
}
\end{lstlisting}
Obviously this corresponds to the expected minimal on-shell basis for $\ord(a)$ corrections to the axial-vector $\AV_{\mu}^{a}$ in Wilson QCD
\begin{align}
\left(\AV_\mu^a\right)_1^{(1)} &= \partial_\mu\bar{\Psi}\gamma_5\tau^a\Psi,&
\left(\AV_\mu^a\right)_2^{(1)} &= \bar{\Psi}M\gamma_\mu\gamma_5\tau^a\Psi,&
\left(\AV_\mu^a\right)_3^{(1)} &= \tr(M)\bar{\Psi}\gamma_\mu\gamma_5\tau^a\Psi.
\end{align}
Here we made explicit use of the fact that Wilson QCD with a canonical degenerate mass term realises exact SU(2)${}_\mathrm{V}$ flavour symmetry and hypercubic symmetry allowing us to relax the fixed Lorentz and isospin coefficient.
While somewhat archaic, running the same script for other choices of fixed $\mu$ and $a$ (including updated transformation behaviours) one would of course reach the same conclusion.

\subsection{Directory and file structure}
The directory of the main script for this introductory example \texttt{findBasisAxial.py} defines the root directory.
Obviously the Python package has to be accessible via \texttt{import opbasis}, which requires the installation of the package.\footnote{Alternatively, the \texttt{opbasis}-folder holding all the source files has to be present in any directory listed under \texttt{sys.path}.}
Both the model file and the Python file implementing the transformation properties have to be located in the subdirectories \texttt{models} and \texttt{transformations} of the current working directory respectively, here their relative paths are \texttt{models/axial.in} and \texttt{transformations/axial.py}.
The benefit of this approach is that one may easily use multiple models with the same transformations and main script or vice-versa.
This is particularly advantageous if various local fields are targeted with different transformation properties while working within the same theory.

\section{Advanced features}\label{sec:examples}
The primary goal of allowing for the custom implementation of \Block{} aims at the middle piece of \Bilin{}, i.e., everything between the covariant derivatives (or fermion EOMs) acting on the left and right flavour.
This part is accessible via \texttt{Bilinear.blocks} and holds a \texttt{list} of instances of \Block{}.
Different implementations of \Block{} are expected to commute, implicitly assuming that they act in different spaces like colour and spinor.
Meanwhile, the treatment of multiple instances \texttt{b1} and \texttt{b2} of the same \Block{} implementation depends on whether \texttt{isinstance(b1, Multiplicative)} is true.
If so, \texttt{b1} and \texttt{b2} are multiplied together assuming that \texttt{b1.\_\_mul\_\_} has been implemented properly.
Otherwise \texttt{b1} and \texttt{b2} are solely grouped together keeping the ordering intact during simplification.
The custom implementation of \Block{} leaves a lot of freedom to the user which extension to the default behaviour is accessible by implementing \Block{}-specific transformation properties.
For finding templates involving custom implementations of \Block{} with the appropriate overall mass-dimension, one has to assign a non-negative mass-dimension \texttt{md} either implicitly inherited from the parent \Block{} or through explicit use of the decorator \texttt{@massDim(md)}.
The introduction of custom indices derived from \IntIndex{} or \Cindex{} allows control over the range of indices accessible within each index while allowing the user near-complete freedom what to do with those additional indices.
For example, such indices can be used to denote generators acting in flavour space --- for now only SU(2) has been implemented as we have seen earlier.
To summarize, in all those cases the implementation of a custom \Block{} amounts to 
\begin{itemize}
\item For any new implementation of \Block{} carrying indices the use of the decorator \texttt{@indices(suffix,**kwargs)} is mandatory, e.g.,
\begin{lstlisting}[language=Python]
import opbasis as opb
from fractions import Fraction
testIdx = opb.CustomIndex("testIdx", ["a", "b"])
@opb.indices("(%s;%s)", mu = opb.sptIdx, nu = testIdx)
class MyBlock(opb.Block):
   def __init__(self,mu:opb.sptIdx,nu:testIdx,
                factor:int|Fraction|opb.Complex=1):
      self.factor = factor
      self.mu = mu
      self.nu = nu
print(str(MyBlock(opb.sptIdx(0),testIdx.a)))
# => "MyBlock(0;a)"
\end{lstlisting}
where \texttt{MyBlock} has two indices, whose string-representation is fixed by the first argument \texttt{suffix} and the two keyword arguments are necessary to indicate the type of each index.
Notice that the order of indices has to match the order of the function arguments in the initialisation for parsing the string representation of the custom \Block{} while their keys must match the names of the corresponding \texttt{class}-attributes.
It is expected that \texttt{factor} is \emph{always} the last argument with default value set to~$1$.
Preferably, the user should stick to \texttt{[\%s,\%s,...]} as \texttt{suffix} but there is some freedom in the separators and delimiters used.
Please refrain from the use of dots as those are omnipresent as separators between different \Block{} and ensure that \texttt{suffix} starts with a non alpha-numeric character such that the \Block{}-identifiers stay unique.
There exists also the option to infer the indices from the function signature via using \texttt{@defaultIndices} as a decorator instead, which then requires the explicit use of type hints as in the example above and names of the arguments matching those of the \texttt{class}-attributes.
\item If multiple indices are present, \texttt{Block.simplify()} can be used to implement any relations among permutations of the indices.
For example, $F_{\mu\nu}$ being antisymmetric allows to always keep only $\mu<\nu$ by flipping the sign of the overall \texttt{Block.factor} when the indices are exchanged.
\item \texttt{Block.variants()} can be adapted to \texttt{yield} only those variants of this \Block{} implementation that are independent at the level of index permutations.
Again, for $F_{\mu\nu}$ only those variants with $\mu<\nu$ are returned.
While not necessary, it may speed up finding the overcomplete basis.
\item If the mass-dimension of the new implementation of \Block{} differs from its parent, the decorator \texttt{@massDim(md)} should be used to indicate the mass-dimension \texttt{md} which is expected to be a non-negative rational number. 
\end{itemize}
For some inspiration, have a look at \texttt{blocks.py} and \texttt{dirac.py} in the Python package but also the more advanced examples presented in the following.
As long as the customisation is restricted to the class-name, indices, transformation properties, and mass-dimensions the package should be able to cope with any such extensions.
Care has to be taken when deviating from the default generation of the unique string-representation, especially to not break the IO part of the package.

\def\lineFlavourEOM{65}
\def\lineEOMonly{24}
\def\lineNoEOMcount{36}
\def\lineShift{6}
\subsection{SymEFT basis for the action of unrooted Staggered quarks}\label{sec:exampleStaggered}
To highlight a straight-forward generalisation to more complicated flavour-symmetries, let us consider unrooted Staggered quarks, also known as Kogut-Susskind fermions~\cite{Kogut:1974ag}.
Possible choices of a \emph{taste}-representation $\taste$ strictly local in spacetime have been discussed, e.g., in~\cite{Kluberg-Stern:1983lmr,Verstegen:1985kt,Daniel:1986zm,Jolicoeur:1986ek}.
Here it is only relevant that those \emph{tastes} play the role of flavours of a four-fold mass-degenerate continuum theory with lattice artifacts severely breaking the continuum $\text{SU}(4)_\mathrm{L}\times\text{SU}(4)_\mathrm{R}$ flavour symmetry thus lifting the mass-degeneracy as well as modifying all of the common spacetime symmetries.
Following along the lines of our previous work on this topic~\cite{Husung:2025nsv} we have the remaining symmetries
\begin{subequations}\label{eq:StaggeredSymmetries}
\begin{itemize}
\item SU($\Nc$) gauge symmetry,
\item U(1)${}_\mathrm{B}$ flavour symmetry,
\item Remnant of chiral symmetry.
Analogously to conventional chiral symmetry we may introduce the shorthands
\begin{align}
\tastebar_\mathrm{R}&=\tastebar\frac{1-\gamma_5\otimes\tau_5}{2}\,,&\taste_\mathrm{R}&=\frac{1+\gamma_5\otimes\tau_5}{2}\taste\,,\nonumber\\\tastebar_\mathrm{L}&=\tastebar\frac{1+\gamma_5\otimes\tau_5}{2}\,,&
\taste_\mathrm{L}&=\frac{1-\gamma_5\otimes\tau_5}{2}\taste\,.
\end{align}
The massless lattice action written in this form is then invariant under
\begin{align}
\tastebar_\mathrm{R}&\rightarrow\tastebar_\mathrm{R}e^{-i\varphi_\mathrm{R}},& \taste_\mathrm{R}&\rightarrow e^{i\varphi_\mathrm{R}}\taste_\mathrm{R},\nonumber\\
\tastebar_\mathrm{L}&\rightarrow\tastebar_\mathrm{L}e^{-i\varphi_\mathrm{L}},&\taste_\mathrm{L}&\rightarrow e^{i\varphi_\mathrm{L}}\taste_\mathrm{L},
&\varphi_\mathrm{L,R}&\in\mathbb{R}.\label{eq:LRsymm}
\end{align}
\item \emph{Modified} charge conjugation
\begin{align}
\tastebar(y)&\rightarrow-\taste^T(y)C\otimes (C^{-1})^T,\quad \taste(y)\rightarrow C^{-1}\otimes C^T\tastebar^T(y),\nonumber\\
U_\mu(x)&\rightarrow U_\mu^*(x),\quad C\gamma_\mu C^{-1}=-\gamma_\mu^T.
\end{align}
\item[$\circ$] \emph{Modified} Euclidean reflections~\cite{Verstegen:1985kt} in $\hat{\mu}$ direction
\begin{align}
\tastebar(y)&\rightarrow \tastebar(y-2y_\mu\hat{\mu})\gamma_5\gamma_\mu\otimes \tau_5\left\{1+a^2\ldots\right\},\nonumber\\
\taste(y)&\rightarrow\gamma_\mu\gamma_5\otimes\tau_5\left\{1+a^2\ldots\right\}\taste(y-2y_\mu\hat{\mu})\,,\nonumber\\
U(x,\nu)&\rightarrow\begin{cases}
U^\dagger(x-(2x_\mu+a)\hat{\mu},\mu) & \mu=\nu \\
U(x-2x_\mu \hat{\mu},\mu)U(x-(2x_\mu-a)\hat{\mu},\nu)U^\dagger(x-2x_\mu\hat{\mu}+a\hat{\nu},\mu) & \text{else}
\end{cases}.\label{eq:modReflections}
\end{align}
\item[$\circ$] \emph{Modified} discrete rotations~\cite{Mitra:1983bi,Verstegen:1985kt} of $90^\circ$ in any $\rho$-$\sigma$-plane, i.e., $\rho<\sigma$,
\begin{align}
\tastebar(y)&\rightarrow\frac{1}{2}\tastebar(R^{-1}y)(\unity +\gamma_\rho\gamma_\sigma)\otimes (\tau_\rho-\tau_\sigma)\{1+a^2\ldots\},\nonumber\\
\taste(y)&\rightarrow\frac{1}{2}(\unity -\gamma_\rho\gamma_\sigma)\otimes (\tau_\rho-\tau_\sigma)\{1+a^2\ldots\}\taste(R^{-1}y),\nonumber\\
U(x,\nu)&\rightarrow\begin{cases}
U^\dagger(R^{-1}x-a\hat{\sigma},\sigma) & \nu=\rho\\
U(R^{-1}x,\sigma)U(R^{-1}x+a\hat{\sigma},\rho)U^\dagger(R^{-1}x+a\hat{\rho},\sigma) & \nu=\sigma\\
U(R^{-1}x,\sigma)U(R^{-1}x+a\hat{\sigma},\nu)U^\dagger(R^{-1}x+a\hat{\nu},\sigma) & \text{else}
\end{cases}\nonumber\\
(R^{-1}x)_\rho&=x_\sigma,\quad (R^{-1}x)_\sigma=-x_\rho,\quad (R^{-1}x)_{\mu\neq \rho,\sigma}=x_\mu,
\end{align}
with rotation matrix $R$ acting on vectors in Euclidean spacetime.
\item[$\circ$] Shift-symmetry by a single lattice spacing in direction $\hat{\mu}$ combined with a discrete flavour rotation and field redefinition
\begin{align}
\bar\taste(y)&\rightarrow \bar\taste(y)\unity\otimes\tau_\mu \left\{1+2a\hat{P}_-^{(\mu)}\hat{\overline{\nabla}}{}_\mu^{\dagger}+a^2\ldots\right\},\nonumber\\
\taste(y)&\rightarrow \unity\otimes\tau_\mu \left\{1+2a\hat{P}_+^{(\mu)}\hat{\overline{\nabla}}_\mu+a^2\ldots\right\}\taste(y),\nonumber\\
U(x,\nu) &\rightarrow U(x,\mu)U(x+a\hat{\mu},\nu)U^\dagger(x+a\hat{\nu},\mu),
\end{align}
where $\hat{\overline{\nabla}}_\mu$ involves a $\hat\mu$-dependent fat link and we introduced the projector
\begin{equation}
\hat{P}_\pm^{(\mu)}=\frac{1\pm\gamma_\mu\gamma_5\otimes\tau_\mu\tau_5}{2}.
\end{equation}
\end{itemize}
\end{subequations}
Filled ($\bullet$) or open ($\circ$) symbols highlight symmetries that are exact counterparts of their continuum-theory variant or require field-redefinitions respectively.
For compactness we only wrote out terms in those field-redefinitions in the lattice theory with canonical mass-dimension up to 1.
For a more in-depth discussion of the implications of those field-redefinitions, see~\cite{Husung:2025nsv}.
As it turns out, the need for such field-redefinitions in the lattice theory allows for the presence of EOM-vanishing operators in the SymEFT action that are incompatible with the leading order part of the symmetry.
Such EOM-vanishing operators are to leading order in the lattice spacing irrelevant for spectral quantities but may have an impact through contact terms on matrix elements~\cite{Capitani:1999ay,Capitani:2000xi,Husung:2024cgc} or, at higher orders in the lattice spacing, on spectral quantities as well.
By symmetry arguments, one can show that only Shift-symmetry may be broken in the SymEFT at $\ord(a)$ by EOM-vanishing operators.

The symmetry transformations from \eqs{eq:StaggeredSymmetries} can be easily mapped into a model file, listing~\ref{code:StaggeredModel}, and a Python module \texttt{Staggered.py} provided in the supplemental material.
The latter also implements the custom \Block{} implementation \texttt{Taste} see listing~\ref{code:StaggeredModule}, which amounts to Dirac Gamma matrices acting in flavour (or ``taste'') space with slight modifications to their behaviour under Euclidean reflections and rotations.\footnote{Here we choose to implement the transformations directly at the level of Dirac Gamma and Taste matrices for simplicity. It would be more natural to implement these transformations at the level of the \Bilin{}.}
Contrary to Wilson quarks, Staggered quarks realise a remnant of chiral symmetry, which here is being implemented as a \texttt{Spurion}.
The concept of a \texttt{Spurion} at the level of this package is to veto operators that do not comply with an assumed spurionic symmetry, here chiral symmetry with masses promoted to Spurion fields.
It is instructive to run this setup for operators of mass-dimension 3~to~5 and indeed find a minimal basis containing only the operators of mass-dimension~4 expected for continuum QCD with four mass-degenerate quarks --- thus confirming once more the absence of additive renormalisation for the action and $\ord(a)$ improvement for spectral quantities in this lattice setup.
The full workflow is contained in \texttt{findBasisStaggered.py} as part of the supplemental material.
Due to the presence of flavour-space matrices we have to introduce the identity matrix \texttt{Taste[id\_]} in the bilinear of the gluon EOM in line~\lineFlavourEOM{}.
Alternatively it should be possible to drop any occurrence of \texttt{Taste[id\_]}.
At mass-dimension~6 and thus at $\ord(a^2)$ we find precisely the minimal on-shell operator basis stated in~\cite{Husung:2025nsv}.
Note that working out this basis is rather involved and may take a couple of minutes.

The unusual symmetries \eqs{eq:StaggeredSymmetries} realised by unrooted Staggered quarks allows for EOM-vanishing operators undergoing less-strict symmetry constraints.
To check for such operators at $\ord(a)$ we comment out line~\lineShift{} from listing~\ref{code:StaggeredModel} and thus remove Shift-symmetry as a constraining symmetry.
Instead we uncomment line~\lineEOMonly{} and comment out line~\lineNoEOMcount{} in \texttt{findBasisStaggered.py} to only allow for templates of operators vanishing by the EOMs and prioritise explicitly mass-dependent operators.
Those minimal changes then lead to the minimal EOM-vanishing basis for the unrooted Staggered quark action at $\ord(a)$
\begin{align}
\mathcal{O}_{\mathcal{E};1}^{(1)} &= \frac{1}{2}\sum_\nu\bar\Psi\left\{\cev{\slashed{D}}\cev{D}_\nu \gamma_\nu\gamma_5\otimes\tau_\nu+\gamma_\nu\gamma_5\otimes\tau_\nu D_\nu\slashed{D}\right\}\Psi,\nonumber\\
\mathcal{O}_{\mathcal{E};2}^{(1)} &= \frac{1}{2}\sum_\nu\bar\Psi\left\{\cev{\slashed{D}}M \gamma_5\otimes\tau_\nu+\gamma_5\otimes\tau_\nu M\slashed{D}\right\}\Psi,
\end{align}
where $\slashed{D}=\gamma_\mu D_\mu + M$ and $\cev{\slashed{D}}=\gamma_\mu \cev{D}_\mu - M$.

\begin{figure}
\begin{lstlisting}[language=modelFile,caption={Model file declaring symmetry transformations and custom \Block{} for unrooted Staggered quarks.},label=code:StaggeredModel]
Transf: Staggered
Flavours: u1 u2 u3 u4 = Psi
Block: Taste
Discrete: modP ++++
Discrete: modC +
Discrete: Shift ++++
Discrete: modH4 ++++++
Spurion: remnantChiral
\end{lstlisting}
\end{figure}
\begin{figure}
\begin{lstlisting}[language=Python,caption={Implementation of \texttt{Taste}-space matrices for unrooted Staggered quarks. All modifications to rotations, reflections and chiral Spurion transformation are implemented as part of \texttt{Taste}.},lastline=16,label=code:StaggeredModule]
import opbasis as opb
@opb.defaultIndices
class Taste(opb.Gamma):
   def reflection(self, mu:int):
      _t5 = self.__class__(opb.Dirac.g5)
      return _t5 * self * _t5
   def rotation(self, plane:int):
      rho,sigma = opb.rotPlanes[plane]
      trho = self.__class__(opb.axisGammas[rho])
      tsigma = self.__class__(opb.axisGammas[sigma])
      _t5 = self.__class__(opb.Dirac.g5)
      test = trho*self*trho
      test2 = tsigma*self*tsigma
      if test == test2:
         return _t5*test*_t5
      return -_t5*trho*self*tsigma*_t5
\end{lstlisting}
\end{figure}

\subsection{\boldmath $B^*\pi$ excited-state contamination for $B$ meson}\label{sec:exampleBstarPi}
To also give an example beyond SymEFT (or generally beyond EFTs), we will derive here the appropriate interpolating operator to couple predominantly to $B^*\pi$ excited-state contaminations affecting matrix elements of asymptotic $B$-meson states, see~\cite{Bar:2023sef}.
While such an interpolator can also be obtained straight-forwardly with pen-and-paper or by the use of character tables to project into the appropriate irreducible representation, we use this example to show-case a few more useful features of this package.

To further simplify the problem at hand we will forget about the notion of isospin and label all light quarks as \texttt{l}.
In practice one would need to further project into the isospin doublet through proper combinations of up and down quarks.
As before we need to specify the various transformation properties of relevance, here of the vector $B^*$ listing~\ref{code:BstarModel} as well as the pseudo-scalar pion and $B$ meson listing~\ref{code:PseudoModel}.
\begin{figure}\centering
\captionof{lstlisting}{Model files describing the spatial transformation properties of the desired meson interpolators.}
\begin{sublstlisting}{0.49\textwidth}
\caption{Vector}\label{code:BstarModel}
\begin{lstlisting}[language=modelFile]
Transf: mesons
Flavours: l b = Psi
Discrete: P x-++
Discrete: H4 xxxxx+
\end{lstlisting}
\end{sublstlisting}
\begin{sublstlisting}{0.49\textwidth}
\caption{Pseudo-scalar}\label{code:PseudoModel}
\begin{lstlisting}[language=modelFile]
Transf: mesons
Flavours: l b = Psi
Discrete: P x---
Discrete: H4 xxx+++
Block: Boost
\end{lstlisting}
\end{sublstlisting}
\end{figure}
For the 2-particle interpolator we further need to introduce the notion of a momentum, here via the implementation of \texttt{Boost} that can be found in listing~\ref{code:Boost} while the full Python module \texttt{mesons.py} can be found in the supplemental material.
\begin{figure}
\begin{lstlisting}[language=Python,caption={Implementation of \texttt{Boost} and custom indices to represent injected momenta.},firstline=5,firstnumber=5,lastline=26,label=code:Boost]
momIdx = opb.IntIndex("momIdx", names=["m1","null","p1"], start = -1)
@opb.defaultIndices
class Boost(opb.Block):
   def __init__(self, px:momIdx, py:momIdx, pz:momIdx,
                factor:int|Fraction|opb.Complex=1):
      self.px = px
      self.py = py
      self.pz = pz
      self.factor = factor
   def rotation(self, plane:int):
      rho,sigma = opb.rotPlanes[plane]
      temp = [x.value for x in (self.px,self.py,self.pz)]
      temp2 = _copy(temp)
      temp2[rho-1] = temp[sigma-1]
      temp2[sigma-1] = -temp[rho-1]
      return self.__class__(*map(momIdx, temp2), self.factor)
   def reflection(self, mu:int):
      return self.__class__(
         momIdx(-self.px.value if mu==1 else self.px.value),
         momIdx(-self.py.value if mu==2 else self.py.value),
         momIdx(-self.pz.value if mu==3 else self.pz.value),
         self.factor)
\end{lstlisting}
\end{figure}
Since we now exactly what kind of \texttt{template} to use here we can directly input \texttt{b.Gamma.l} and \texttt{l.Gamma.l} as argument when calling \texttt{overcompleteBasis(template, model)} for the $B^*$ and pion respectively.
This produces the expected bilinears $\bar{\mathrm{b}}\gamma_1\mathrm{l}$ and $\bar{\mathrm{l}}\gamma_5\mathrm{l}$ respectively as well as their counter-parts with an additional $\gamma_0$ inserted since we are only restricting spatial quantum numbers here.

To obtain an ansatz for a pseudo-scalar $B^*\pi$ interpolator at rest we insert appropriate \texttt{Boost} into the $B^*$-like and pion-like bilinear and then multiply them together.
After a call to \texttt{symmetrise(bs*pi, model)} we then obtain the desired pseudo-scalar interpolators such as
\begin{lstlisting}[numbers=none,xleftmargin=0em,framexleftmargin=0em,tabsize=3]
+1*{
	+1*<b.Gamma[gx].Boost[-1,0,0].l l.Gamma[g5].Boost[1,0,0].l>
	-1*<b.Gamma[gx].Boost[1,0,0].l l.Gamma[g5].Boost[-1,0,0].l>
	+1*<b.Gamma[gy].Boost[0,-1,0].l l.Gamma[g5].Boost[0,1,0].l>
	-1*<b.Gamma[gy].Boost[0,1,0].l l.Gamma[g5].Boost[0,-1,0].l>
	+1*<b.Gamma[gz].Boost[0,0,-1].l l.Gamma[g5].Boost[0,0,1].l>
	-1*<b.Gamma[gz].Boost[0,0,1].l l.Gamma[g5].Boost[0,0,-1].l>
}
\end{lstlisting}
Notice that in each step the appropriate model has to be used to have appropriate transformation properties and allow the use of \texttt{Boost}.
The full workflow can be found in the supplemental material as \texttt{findIrrep.py}.

\section{Summary}
The combination of defining custom implementations of \Block{} together with the ability to specify custom transformation properties allows access to a wide range of models and cases --- probably even various ones that have not been mentioned or thought of here.
Originally thought of as a tool for SymEFT, its applicability should be open to other EFTs, (brute-force) derivation of irreducible representations as well as finding all operators allowed to mix under renormalisation given (reduced) lattice transformation properties.
This approach then offers relative ease in deriving the desired operator bases.

For SymEFT this is an important step forward for a more systematic analysis of the $\ord(a)$ but more importantly $\ord(a^2)$ lattice artifacts of various local fields as well as other discretisations of the lattice action.
Possible extensions could be, for example the energy momentum tensor, see e.g.~\cite{DallaBrida:2020gux}, or Karsten-Wilczek fermions~\cite{Karsten:1981gd,Wilczek:1987kw} respectively.
Already Wilson-like theories like Wilson quarks with a maximal chiral twist~\cite{Frezzotti:2005gi} or mixed actions have an abundance of use cases to be explored.
Another interesting direction could be addressing new ideas for lattice actions to check for potential additive renormalisations in a more automated fashion.

While the user still needs to work out all the relevant transformation properties, the entire work of finding a minimal basis of operators using EOMs and integration-by-parts relations is being taken care of.
Meanwhile, some level of control is maintained regarding which operators to keep as discussed in \sect{sec:basisReduction} through an appropriate ordering of the initial templates.
In particular, this allows to accommodate for scenarios like lattice actions with non-trivial boundary conditions thus requiring to keep some total derivatives when deriving the SymEFT action.

An obvious generalisation of this package would be to allow for multiple gauge interactions as well as to relax the constraint to have solely traceless generators.
As it turns out, tracking the Abelian part of the theory in the current setup is surprisingly difficult and will require a major overhaul.
After initial attempts this has been disregarded for the moment.
Obviously, as the lattice QCD community now starts to incorporate QED effects, see e.g.~\cite{Carrasco:2015xwa,Lucini:2015hfa}, such a generalisation would be desirable for the future.

\paragraph{Acknowledgements.}
I thank Gregorio Herdoíza for encouraging me to develop this code and make it publicly available as well as for various feedback on the manuscript.
I also thank Hubert Simma for useful discussions on the manuscript, and Fernando Pérez Panadero as well as Javier Carrón Duque for helpful suggestions on the details of the reduction to a minimal operator basis.
The author acknowledges support by the projects PID2021-127526NB-I00, funded by MCIN/\allowbreak AEI/\allowbreak 10.13039/\allowbreak 501100011033 and by FEDER EU, as well as IFT Centro de Excelencia Severo Ochoa No CEX2020-001007-S, funded by MCIN/\allowbreak AEI/\allowbreak 10.13039/\allowbreak 501100011033.

\appendix
\begin{landscape}
\section{Structure of \LC{}, \Bilin{} and \Tr{} implementations}\label{sec:structBilinTrace}
\begin{figure}[h]\centering
\includegraphics[page=4]{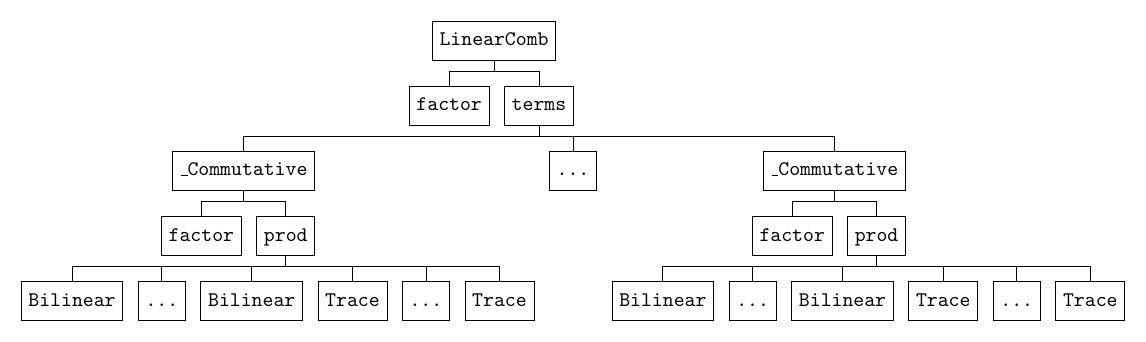}
\caption{Schematic of the internal representation of \LC{} by various commuting terms carrying distinct factors and products of different instances of \Tr{} and / or \Bilin{}.
The latter are sketched in more detail in \figs{fig:Bilin}--\ref{fig:otherTrace}.}
\label{fig:LinearComb}
\end{figure}
\begin{figure}[h]\centering
\includegraphics[page=1]{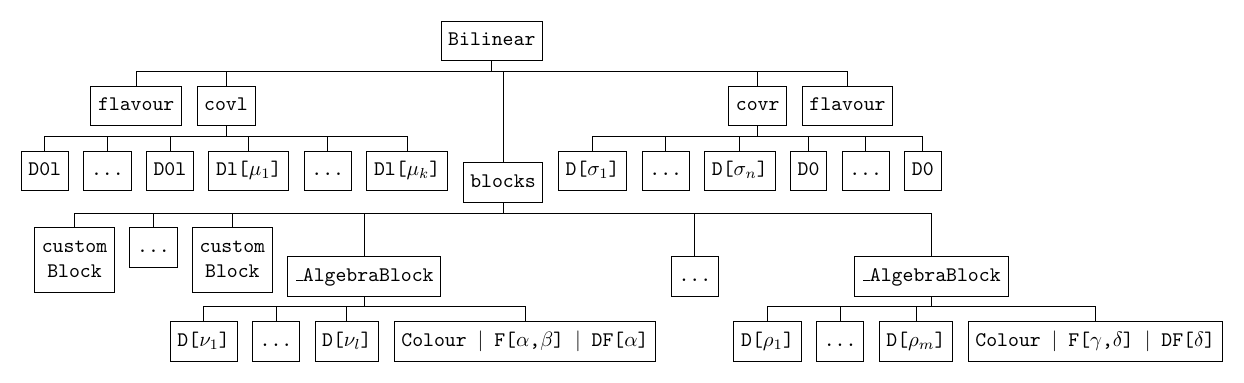}
\caption{Schematic of the three major parts forming a \Bilin{} namely the left and right flavour dressed with covariant derivatives (or fermion EOMs \texttt{D0l} and \texttt{D0}) stored in \texttt{covl} and \texttt{covr} respectively as well as everything in between those flavours stored in \texttt{blocks}.
The main customisability offered by this package is due to the custom implementation of \Block{} that can be stored in \texttt{blocks}.
\AlgBlock{} is a protected implementation taking care of an element of the algebra optionally with covariant derivatives acting on it.
The three possible insertions of elements of the algebra are the dummy generator \texttt{Colour}, the field strength \texttt{F}, and the gluon EOM \texttt{DF}.}
\label{fig:Bilin}
\end{figure}
\begin{figure}[h]\centering
\includegraphics[page=2]{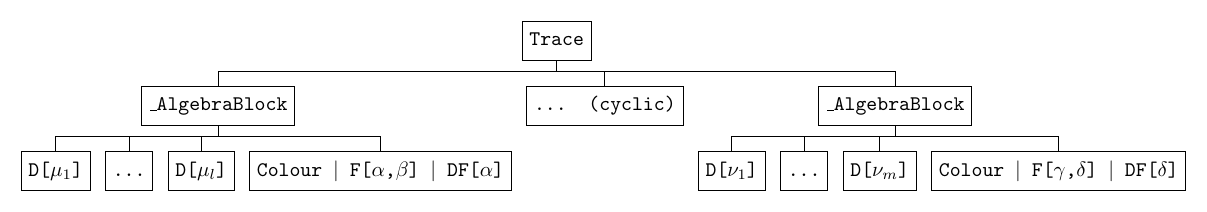}
\caption{Schematic of the algebra \Tr{} which \emph{only} allows elements of \AlgBlock{} to be present.
\AlgBlock{} is a protected implementation taking care of an element of the algebra optionally with covariant derivatives acting on it.
The three possible insertions of elements of the algebra are the dummy generator \texttt{Colour}, the field strength \texttt{F}, and the gluon EOM \texttt{DF}.}
\label{fig:colourTrace}
\end{figure}
\begin{figure}[h]\centering
\includegraphics[page=3]{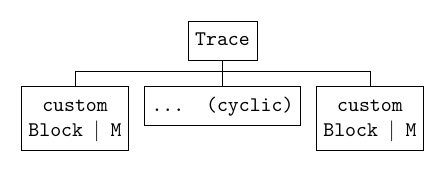}
\caption{Schematic of the \Tr{} acting on mass matrices or other custom implementations of \Block{}.
Only implements cyclicity.
Any further simplification has to be implemented manually.}
\label{fig:otherTrace}
\end{figure}
\end{landscape}

\section{Wilson QCD}\label{sec:Wilson}
The primary example used throughout the paper to highlight common usecases are two mass-degenerate Wilson quarks~\cite{Wilson:1974,Wilson:1975id}, i.e., we consider the lattice fermion action
\begin{align}
S_\mathrm{F}&=a^4\sum_x\bar{\Psi}(x)\hat{D}_\mathrm{W}\Psi(x)\,,\nonumber\\
\hat{D}_\mathrm{W}&=\frac{\gamma_\mu}{2}\left\{\nabla_\mu+\nabla_\mu^*\right\}+m_0+ac_\mathrm{sw}(g_0^2)\frac{i}{4}\sigma_{\mu\nu}\hat{F}_{\mu\nu}-\frac{ar}{2}\nabla_\mu\nabla_\mu^*\,,
\end{align}
where $\hat{F}_{\mu\nu}$ is a lattice discretisation of the field-strength tensor, typically the clover discretisation, $r\in]0,1]$ commonly chosen to be $r=1$, $m_0$ is the bare quark mass\footnote{Due to the Wilson lattice discretisation breaking flavour symmetries in the massless theory down to SU$(\Nf)_\mathrm{V}$ symmetry, the theory is susceptible to additive renormalisation which is here written as $m_\mathrm{cr}$.
The subtracted quark mass then renormalises multiplicatively as $m_\mathrm{R}=Z_m(m_0-m_\mathrm{cr})$.}, and $2\sigma_{\mu\nu}=i[\gamma_\mu,\gamma_\nu]$.
By proper tuning of $c_\mathrm{sw}(g_0^2)=1+\ord(g_0^2)$ one may further achieve non-perturbative Symanzik $\ord(a)$ improvement of the lattice action~\cite{Luscher:1996sc} in the massless limit.

In the limit of degenerate quark masses the symmetries of Wilson QCD with a canonical mass term can be summarised as, see also~\cite{Sheikholeslami:1985ij},
\begin{itemize}
\item Local SU$(\Nc)$ gauge symmetry
\begin{equation}
\bar{\Psi}(x)\rightarrow \bar{\Psi}(x)\Omega^\dagger(x)\,,\quad \Psi(x)\rightarrow \Omega(x)\Psi(x)\,,\quad U(x,\mu)\rightarrow \Omega(x)U(x,\mu)\Omega^\dagger(x+a\hat{\mu})\,.
\end{equation}
\item Charge conjugation
\begin{equation}
\bar{\Psi}\rightarrow -\Psi^TC\,,\quad \Psi\rightarrow C^{-1}\bar{\Psi}^T\,,\quad C\gamma_\mu C^{-1}=-\gamma_\mu^T,\quad U(x,\mu)\rightarrow U^*(x,\mu)\,.
\end{equation}
\item Euclidean reflections in direction $\hat{\mu}$
\begin{align}
\bar{\Psi}(x) &\rightarrow \bar{\Psi}(x-2x_\mu\hat{\mu})\gamma_5\gamma_\mu\,,\quad \Psi(x)\rightarrow \gamma_\mu\gamma_5\Psi(x-2x_\mu\hat{\mu})\,,\nonumber\\
U(x,\nu) &\rightarrow \begin{cases}
U^\dagger(x-(2x_\mu+a)\hat{\mu},\nu)&\mu=\nu\,,\\
U(x-2x_\mu\hat{\mu},\nu) & \text{else.}
\end{cases}
\end{align}
\item Hypercubic rotations
\begin{align}
\bar{\Psi}(y)&\rightarrow\frac{1}{2}\bar{\Psi}(R^{-1}y)(\unity+\gamma_\rho\gamma_\sigma),\quad\Psi(y)\rightarrow\frac{1}{2}(\unity -\gamma_\rho\gamma_\sigma)\Psi(R^{-1}y),\nonumber\\
U(x,\nu)&\rightarrow\begin{cases}
U^\dagger(R^{-1}x-a\hat{\rho},\rho) & \nu=\sigma\\
U(R^{-1}x,\sigma) & \nu=\rho\\
U(R^{-1}x,\nu) & \text{else}
\end{cases}\nonumber\\
(R^{-1}x)_\rho&=x_\sigma,\quad (R^{-1}x)_\sigma=-x_\rho,\quad (R^{-1}x)_{\mu\neq \rho,\sigma}=x_\mu,
\end{align}
\item SU$(\Nf)_\mathrm{V}$ flavour symmetry
\begin{equation}
\bar{\Psi}\rightarrow \bar{\Psi}V^\dagger\,,\quad \Psi\rightarrow V\Psi\,,\quad V\in\text{SU}(\Nf)\,.
\end{equation}
\end{itemize}

\vskip 0.3cm

\noindent

\bibliographystyle{JHEP}

\providecommand{\href}[2]{#2}\begingroup\raggedright\endgroup

\end{document}